# MODELING, IMPLEMENTATION AND PERFORMANCE ANALYSIS OF MOBILITY LOAD BALANCING FOR LTE DOWNLINK DATA TRANSMISSION

Mohamed Escheikh[1] Hana Jouini[1] and Kamel Barkaoui[2]

[1]University of Tunis El Manar, Enit, Sys'Com Box 37, LE BELVEDERE 1002 Tunis, Tunisia
[2]Conservatoire national des arts et métiers Box 2, conté 75003 Paris, France

## ABSTRACT

*We propose in this paper a simulation implementation of Self-Organizing Networks (SON) optimization related to mobility load balancing (MLB) for LTE systems using ns-3 [1]. The implementation is achieved toward two MLB algorithms dynamically adjusting handover (HO) parameters based on the Reference Signal Received Power (RSRP) measurements. Such adjustments are done with respect to loads of both an overloaded cell and its cells' neighbours having enough available resources enabling to achieve load balancing. Numerical investigations through selected key performance indicators (KPIs) of the proposed MLB algorithms when compared with another HO algorithm (already implemented in ns-3) based on A3 event [2] highlight the significant MLB gains provided in terms global network throughput, packet loss rate and the number of successful HO without incurring significant overhead.*

## KEYWORDS

*LTE networks, Mobility load balancing, Handover, Simulation*

## 1. INTRODUCTION

In recent years the demand for traffic in cellular radio networks has evolved vertiginously. In order to cope with this demand, the organization of international standardization 3rd Generation Partnership Project (3GPP) had introduced the new cellular radio system Long Term Evolution (LTE). This latter adopts a simplified all-IP architecture providing spectral efficiency about two to three times higher than that of the 3GPP Release 6 [3]. LTE will also offer up to 100 Mb/s of throughput on the downlink (DL) with a spectral bandwidth of up to 20 MHz. LTE systems use multiplexing and encoding data technique namely Orthogonal Frequency Division Multiple Access (OFDMA) in the radio interface downlink (DL) transmission and the Single Carrier Frequency Division Multiplexing Access (SC-FDMA) in the uplink (UL) transmission.

For LTE networks, the main challenges are to meet users' quality of service (QoS) requirements especially for real-time traffic in terms of throughput, end to end delay. These challenges concern also the satisfaction of cellular radio operator's requirements in terms of radio resource management (RRM), rationalization of operational expenses and optimization of the overall network efficiency. To address these challenges, SON technologies have been introduced in LTE networks from the 3GPP Release 9 specifications [4]. These technologies are designed to achieve a high level of operational performance by automating a number of tasks such as configuration, optimization and healing (repair). This enables also to reduce capital expenditure (CAPEX) and operational expenditure (OPEX) to deal with the inter-operator competition.

We focus in this article on a key component of SON technologies namely self-optimization and especially load balancing in mobile radio LTE networks. In the literature several research studies have been proposed for investigating the load balancing problem between LTE cells and can be classified in two categories:

- In the first category an overloaded cell (hot-spot cell) trying to borrow resources (radio channels) to the least loaded neighbouring cells [3]. Yao Tien Wang et al. proposed in [5] a method that fits into this category and is based on neural networks and fuzzy logic. The proposed method performs a number of capabilities related to learning, optimization, robustness and fault tolerance. This method is used in order to meet effectively the stringent requirements for multimedia traffic in terms of QoS.

- In the second category the overloaded cell tries to transfer the traffic excess to the least loaded neighbouring cells by dynamically adjusting the handover parameters (hysteresis, time to trigger (TTT) ...) or using the cell breathing technique [6]. The principle of such technique is to gradually shrink the cell coverage as the load increases. In the literature, several research studies have addressed this issue. In [7, 8], new power control algorithms have been proposed in order to dynamically adjust the scope of both overloaded and under-loaded cells. The authors in [9] proposed an algorithm to jointly improve HO performance and load balancing (LB) by introducing a co-weighted satisfaction factor. A typical transfer approach to implement load balancing is presented in [10]. Such approach chooses for this purpose as a source cell, the cell that has the highest utilization ratio and the target cell the cell having the lowest utilization ratio of physical resource blocks (PRB). In [11], cell offset (shift) is automatically adjusted according to the load of the source cell and that of the target neighbouring cell. In [12] the authors propose a method of estimating the load after handover completion. This method is based on the SINR prediction and the measure of the user signal quality.

In [13] authors implemented two elementary procedures (EPs) related to load management (LM) function of the X2-application protocol (X2AP) as specified in TS 136.423 [14].in order to implement a MLB based adaptive handover (HO) algorithm.

The remainder of this paper is organized as follows: in section 2 we will briefly recall architectures and SON features in LTE networks. We detail in this same section the load balancing principle by dynamic HO adjustment. In Section 3 we describe the proposed MLB algorithms. Section 4 is devoted to simulation results and numerical investigation. The last section concludes thus paper.

## 2. SON ARCHITECTURES AND FUNCTIONALITIES FOR LTE NETWORKS

### 2.1. SON Architectures

There are three types of SON architectures in cellular radio networks [15] trying, each one, to find a compromise between stability, scalability and agility. Before detailing the principle these architecture,s we propose to recall the definitions for stability, scalability and agility [16]:

- Scalability: Scalability of the solution is its ability to reduce complexity and to maintain its functionality to provide good performance even for large scale systems.

- Stability: a solution is considered stable if it avoids oscillations (e.g. instability means the occurrence of ping-pong phenomenon when performing handovers in cellular radio networks).

- Agility: is also a key feature of SON systems reflecting the system adaptability in its operational environment changes. Thus an agile algorithm should not react too quickly to temporary changes in the system to prevent large fluctuations between its states.

#### 2.1.1. Centralized SON (C-SON) Architecture

In the C-SON architecture (Figure 1.a) algorithms are executed at the network management system (NMS) level. The main advantage of this approach is that the SON algorithms can gather information from all entities of the network into consideration. This means that it is possible to jointly optimize the parameters of all centralized SON functions. The purpose behind is to provide an overall optimization enabling further stability. Stability is especially useful for networks having characteristics which vary relatively slowly. The C-SON architecture also enables to facilitate coordination between the SON functions. In contrast, C-SON disadvantages are particularly manifested by slow response times, high backbone traffic and a singular point of failure. Notice also that the long duration of response time and the lack of agility can significantly affect the network adaptation rate and cause instability problems.

#### 2.1.2. Distributed SON (D-SON) Architecture

Algorithms in D-SON architectures (Figure 1.b) are executed in the network nodes. Thus the exchange of SON messages can be directly achieved between ENodeBs LTE. Unlike the C-SON architecture, D-SON architecture provides more dynamic for SON features and adapts more quickly to changes in the network (agility). Another advantage of the D-SON solution is its scalability. However, optimizations performed in cells do not necessarily lead to global optimization which may cause undesirable instabilities.

#### 2.1.3. Hybrid SON (H-SON) Architecture

The H-SON architecture (Figure 1.c) executes SON algorithms both in the NMS at the level of network elements (NE). This type of architecture tries to take advantage of the previous two architectures while circumventing disadvantages, which is not always easy to achieve.

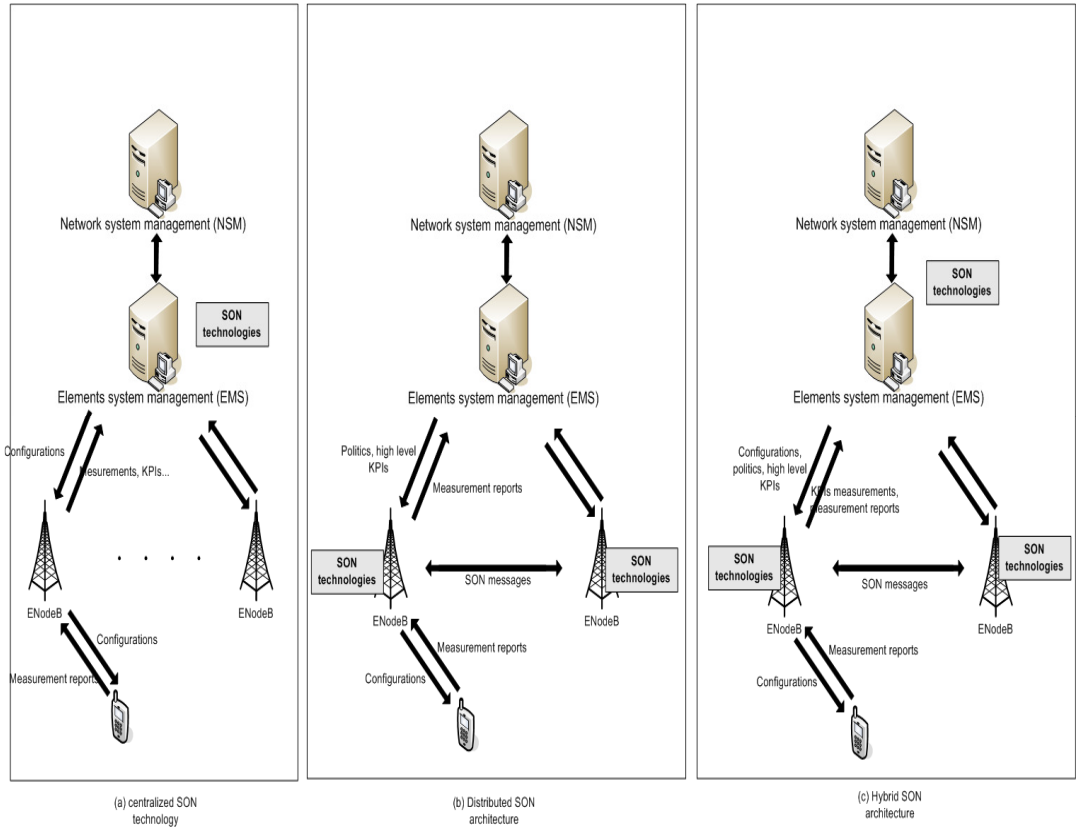

Figure. 1: SON Architectures for LTE networks

## 2.2. SON Features

SON Features in LTE networks include self-configuration, self-diagnosis, self-healing and self-optimizing. The SON technologies are designed automatically to adjust network settings. This is achieved based on measurement reports from UEs and eNodeBs in order to ensure better network service quality and to perform better services for UEs.

### 2.2.1. SON Auto-optimization

Self-optimization is a collection of algorithms aiming to maintain network quality and network performance with minimal human intervention. The self-optimization functions automatically and dynamically trigger, if necessary, optimization actions on the affected NEs. Among the most important functions of self-optimization we distinguished MLB optimization, mobility robustness optimization (MRO). MRO is used in conjunction with MLB to ensure more stability and to mitigate the ping pong phenomenon. Adjusting the settings of MLB and the MRO may be subject to conflicting objectives often harmonized through finding convenient compromise. In this article we focus on MLB.

## 2.3. MLB In LTE Networks Based On Dynamic Adjustment Of HO Parameters

Before detailing the principle of MLB by dynamically adjusting HO parameters, we present the inter-cellular transfer (HO) principle in LTE networks.

### 2.3.1. Handover in LTE Networks

The handover is a major key procedure of mobile cellular radio networks to ensure that users are moving freely across the network while maintaining continuous connectivity and access to services [17]. Given that the handover success rate is a key indicator of network performance, it is essential to achieve a quick and efficient HO process. The optimization of the handover aims to dynamically adjust its setting parameters (offset, hysteresis ...). In LTE, several triggering events may be considered [18]. In this article, we focused on the A3 event. This latter requires that the difference between the received signal level of the current cell and that of the neighbouring cell exceeds a given threshold.

### 2.3.2. MLB In LTE Networks

In LTE networks, the traffic demand of some cells may be much higher than the acceptable level, while other neighbouring cells (of the overloaded cells) may have enough available resources to serve more users. This situation causes, in absence of cooperation, load unbalance between cells.
In order to trigger load balancing between two cells (an overloaded source cell A (hot-spot) and an under-loaded neighbour destination cell B) (Figure 2), two conditions must be satisfied:

- The source cell A load exceeds the predefined threshold to trigger MLB.
- The neighbouring cell B have enough available resources to accept cooperating and handle the cell A traffic excess.

Once the above conditions are met, the cell A chooses, in a first step, a selected number of attached users candidate to achieve HO to cell B. In a next step users adjust their own parameters (handover, reselection, etc.) corresponding to the cell B. In the next section, we describe the principle of load balancing algorithms in mobile networks (MLB) we proposed and implemented in ns-3.

## 3. PROPOSED MLB ALGORITHMS

The two proposed MLB algorithms (*Alg_MLB1* and *Alg_MLB2*) are based on dynamic adaptation of HO parameters (hysteresis values) (Figure 2). The HO due to MLB is the transition of the current overloaded cell (cell A) to one or more neighbouring cell(s) willing to cooperate. The performance of these two algorithms will be compared to a standard HO algorithm without MLB (*Alg_without_MLB*). Note that the above algorithms (with and without MLB) are based on measuring the RSRP and use the A3 event of the 3GPP specification [2] for triggering the HO. The origin of the HO in LTE networks implementing the MLB can be of two different types. The first type is the realization of the A3 event, while the second refers to a load mismatch between two neighbouring cells. Before detailing the principle of the two MLB algorithms, we will in what follows set some notations and preliminary thresholds that will be commonly used in the description of both algorithms *Alg_MLB1* and *Alg_MLB2*.

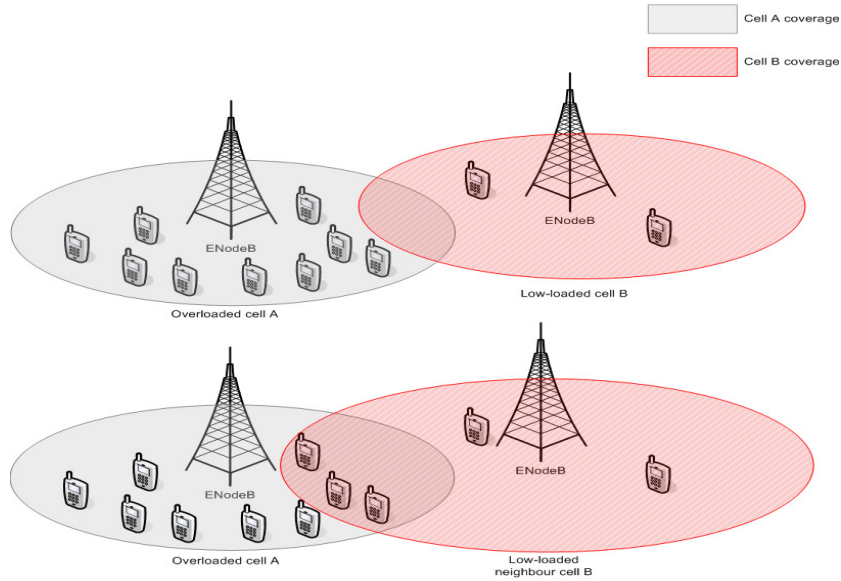

Fig. 2: MLB with dynamic HO adjustment

**Preliminary Notations:**

- $V_{AR}(i)$: (Available Resources) The available amount of resources in $i^{th}$ cell.
- $V_{TR}(i)$ (Total Resources): The total amount of resources in $i^{th}$ cell. Notice for conventional notation that $i = 0$ indicates the current cell and $i > 0$ indicates the $i^{th}$ neighbouring cell.
- $V_{AR}(i)/V_{TR}(i)$: The relative amount (in %) of available resources in cell $i$.
- $Th_{PreMLB}$: The predefined threshold for triggering MLB (Figure 3).
- $Th_{postMLB}$: The threshold for disabling MLB.
- $Th_{AvailMLB}$: The threshold to accept the MLB request.

The current cell is assumed overloaded if the following condition is satisfied:

$$\frac{V_{AR}(0)}{V_{TR}(0)} < Th_{Pr\,eLB} \qquad (condition\ 1)$$

If *condition 1* is verified, the MLB procedure is triggered. The overloaded cell dynamically configures the new HO hysteresis thresholds for different neighbouring cells with respect to their relative amount of available resources. These new (updated) thresholds are calculated from the following equation [19]:

$$Th_{Hys}(0,i) = \alpha_i . Th_{Hys}(0), \quad (0 \leq \alpha_i \leq 1) \tag{1}$$

with and

$$\alpha_i = \begin{cases} 0 & if \ \frac{V_{AR}(i)}{V_{TR}(i)} > Th_{PostMLB} \\ \beta_i & if \ Th_{AvailMLB} \leq \frac{V_{AR}(i)}{V_{TR}(i)} \leq Th_{PreMLB} \\ 1 & if \ \frac{V_{AR}(i)}{V_{TR}(i)} < Th_{AvailMLB} \end{cases} \quad (2)$$

$$\beta_i = \begin{cases} 1 - \frac{Th_{AvailMLB} - \frac{V_{AR}(i)}{V_{TR}(i)}}{Th_{AvailMLB} - Th_{PostMLB}} & for \ Alg\_MLB1 \\ 0,5 & for \ Alg\_MLB2 \end{cases} \quad (3)$$

Notice that $Th_{Hys}(0)$ is the HO hysteresis threshold of the current cell before triggering MLB and $Th_{Hys}(0, i)$ the HO hysteresis threshold of the current cell (with index 0) toward the neighbouring cell with index $i$.

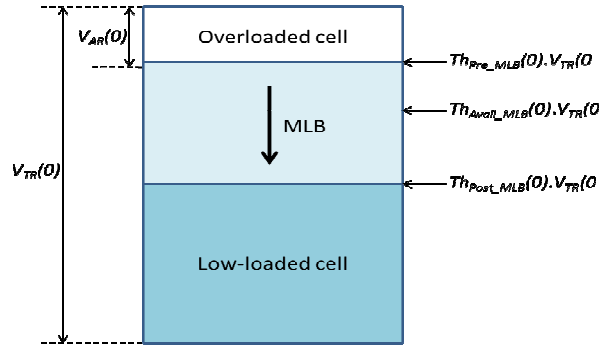

Fig. 3: Triggering condition of the MLB algorithms in the current cell

As soon as MLB condition is satisfied (i.e. *condition 1*), the current cell updates HO hysteresis threshold $Th_{Hys}(0,i)$ and sends then the active UEs (attached to current cell) via a control message measurement. In turn, the UEs update the new hysteresis thresholds as long as the condition on the A3 event is checked. Moreover, once the *condition 2* is verified, the MLB is disabled.

$$\frac{V_{AR}(0)}{V_{TR}(0)} > Th_{PostLB} \qquad (condition\ 2)$$

From the above description, we notice that the MLB is effective unless the two conditions (*condition 1* for the current cell and *condition 3* for the neighbouring cell) are verified:

$$\frac{V_{AR}(i)}{V_{TR}(i)} > Th_{AvailLB} \qquad (condition\ 3)$$

Thus, one can distinguish an alternation between periods with and without MLB depending on load variations of both the current cell and the neighbouring cell(s). Note that as soon as the deactivation condition of the MLB process (*condition 2*) holds, the current cell needs to recover the original hysteresis value by sending another measurement control message to the UEs. Figure 4 depicts the principle of the proposed MLB algorithms.

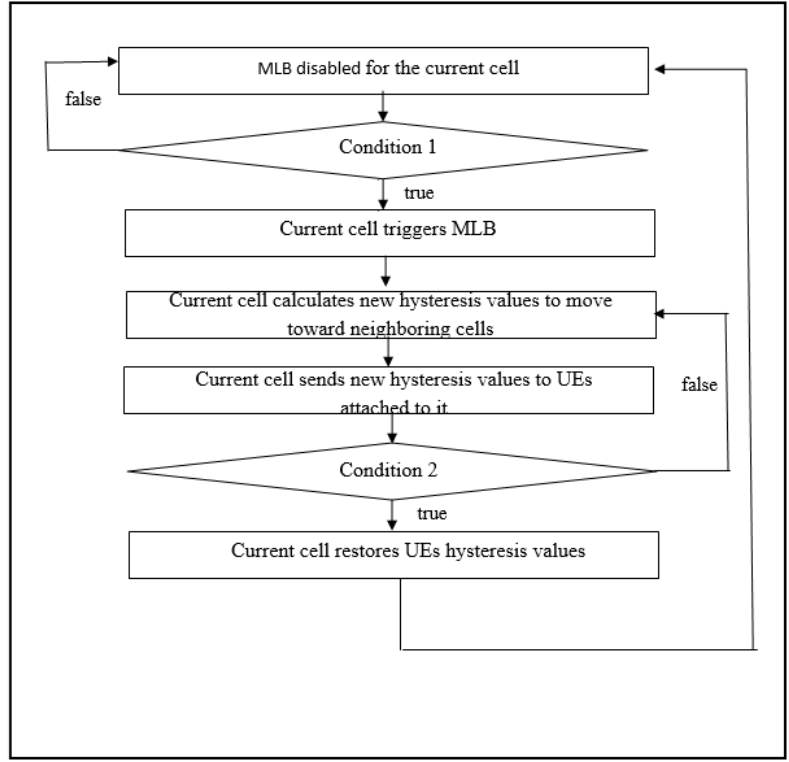

Fig. 4: Principle of the proposed MLB algorithms implemented in ns-3

## 4. SIMULATION ANALYSIS AND NUMERICAL RESULTS

To highlight the performance of the proposed MLB algorithms (i.e. *Alg_MLB1* and *Alg_MLB2*), we compare their performances with those of a standard algorithm without MLB (*Alg_without_MLB*) for different UEs densities (i.e. for different global loads). *Alg_without_MLB* only implements the A3 event of the 3GPP specification and without accounting of load balancing. Such algorithm is used in our case study as a standard algorithm. *Alg_MLB1* and *Alg_MLB2*, associated with the A3 event, aim to ensure better distribution of the overall LTE network load and to reduce packet losses.

### 4.1. Simulation Model

We design a simulation model composed of three equi-spaced eNodeBs with distance of 500m. Each eNodeB includes three sectors. MLB algorithms are implemented in a distributed manner (D-SON) in each sector. These algorithms dynamically adjust the HO parameters (hysteresis and time to trigger (TTT)) as a function of the network load. TTT is a timer used in the 3GPP specification to provide greater robustness and to better mitigate the ping-pong phenomenon. In the investigated scenarios, we have focused on the dynamic adjustment of the hysteresis in the range between 0 db and 3 db in steps of 0.5 dB for fixed values of TTT. Table 1 summarizes the correspondence between the modulation and coding schemes (MCS) (used in our simulation model) and the total capacity of a cell in Mbps for a bandwidth of 5MHz (25RBs).

Table 1: Correspondence between MCS and the total cell capacity in Mbps for bandwidth of 5Mhz (25RBs)

| MCS | Modulation | Total cell capacity (in Mbps) |
|---|---|---|
| [0..9] | QPSK | 13.2 |
| [10..16] | 16QAM | 26.4 |
| [17..28] | 64QAM | 39.6 |

Table 2 recapitulates the main simulation parameters used in our case study.

TABLE 2: Simulation parameters

| Parameter | Value |
|---|---|
| Simulation duration | 100 s |
| Number of eNodeBs | 3 (9 sectors) |
| Distance between eNodeBs | 500 m |
| Power transmitted by a eNodeB in DL | 46 dBm |
| Traffic nature | TCP |
| UEs density (37/56/75 UEs) | 2/6/8 * E-05 |
| Minimal UE moving speed (60 Km/h) | 16.6667 m/s |
| Maximal UE moving speed (60 Km/h) | 16.6667 m/s |
| Bandwidth in UL and DL | 5 MHz (25 RBs) |
| Time To Trigger | 256 ms |
| Default hysteresis value | 3 dB |
| Hysteresis margin with MLB | [0..3 dB] |
| $Th_{PreMLB}$ | 0.2 |
| $Th_{AvailMLB}$ | 0.3 |
| $Th_{PostMLB}$ | 0.4 |

In the next section, we present numerical results related to simulation investigations of the proposed scenario.

### 4.2. Numerical Results

Figure 5 shows that the global network throughput provided by *Alg_MLB1* and *Alg_MLB2* is greater than that provided by *Alg_without_MLB* for different UEs densities (37, 56, 75 UEs). This may be justified by the MLB principle which favours the transfer of a part of the exceeded load from the congested cell to the under-loaded neighbouring cells. Thus, the risk of traffic loss in overloaded cell will be reduced, and possibly avoided by evenly spreading this load excess toward under-loaded neighbouring cells. Another interesting observation can be done from Figure 5 is that the MLB certainly improves the global network throughput for different loads but this improvement is not very significant for high load. The explanation of such behaviour is that for such load it's very likely that neighbouring cells (of an overloaded cell requesting their cooperation for possible load balancing) are themselves overloaded. This intuitively corresponds to a *condition 2* highly probably not verified. We have also investigated the average throughput per sector in DL for different UEs loads (37, 56 and 75 UEs). It is worth noting that the MLB implementation improves the throughput of most areas (2, 4, 7, 8 and 9) (Figure 6). The results for the sectors 1, 3, 5 and 6 are mitigated. This means that one of the two proposed algorithms (*Alg_MLB1* and *Alg_MLB2*) provides a better throughput than *Alg_without_MLB*.

Thanks to MLB implementation traffic in overloaded cells is substantially decreased since the

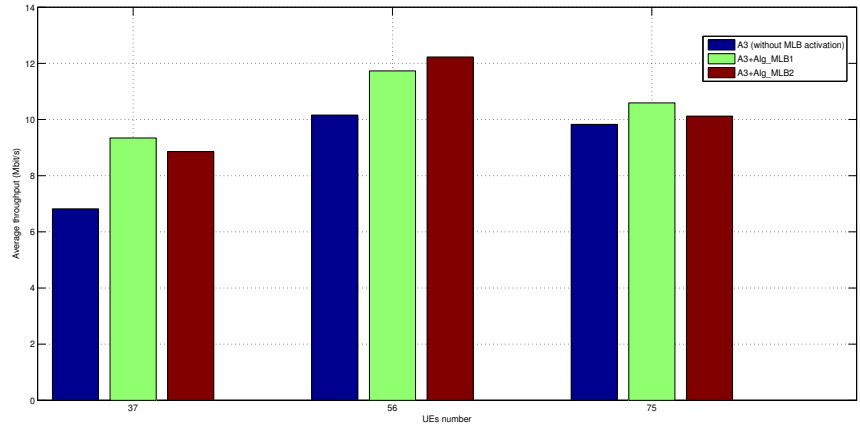

Figure. 5: Global network throughput in DL with respect to UEs density

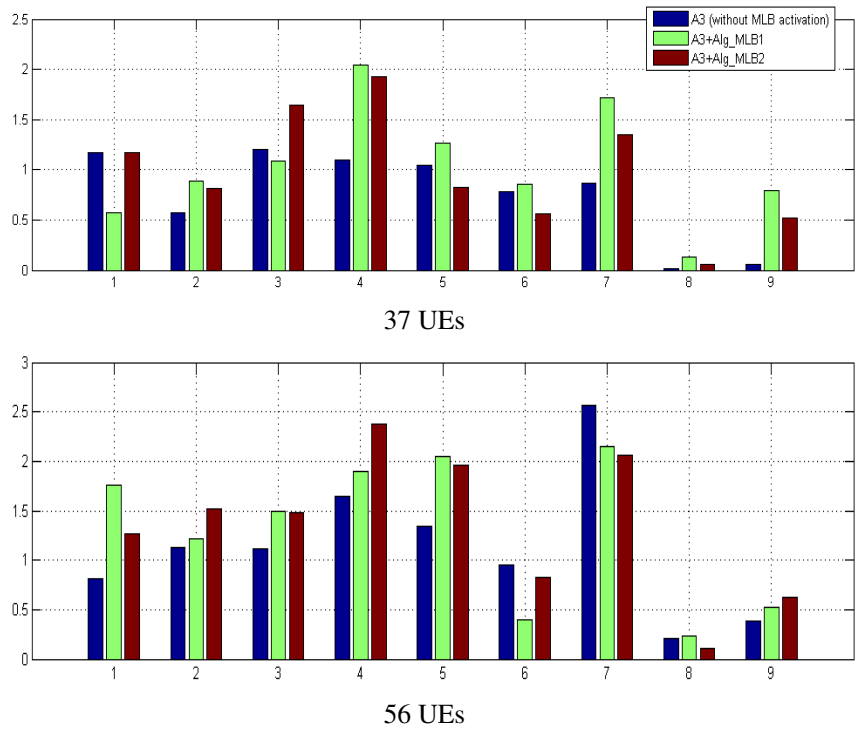

37 UEs

56 UEs

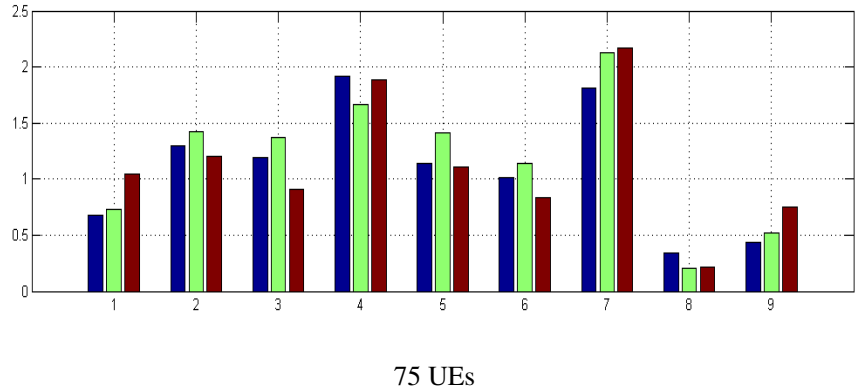

75 UEs

Figure. 6: Average throughput per sector in DL for different UEs densities (*Alg_MLB1 Alg_MLB2 and Alg_without_MLB*)

exceeded load is transferred to the less loaded neighbouring cell(s). This results in a significant reduction of the packet loss ratio compared to *Alg_without_MLB* (Figure 7). From Figure 7, the loss ratio is an increasing function of the network load. With MLB, losses may occur in a cell if it is enough overloaded and cannot find a neighbouring cell willing to cooperate. In other words, the cell loss ratio of the current cell closely depends on both its own load and the load of the neighbouring cell(s). Since MLB aims to reduce the loss ratio through an evenly load distribution between cells, we observe a significant reduction of this metric for the scenarios with MLB when compared with the scenario without MLB (Figure 7). Figure 8 shows the evolution of the number of successful HO with respect the global network load (UEs density for the three investigated algorithms (*Alg_MLB1, Alg_MLB2* and *Alg_without_MLB*).

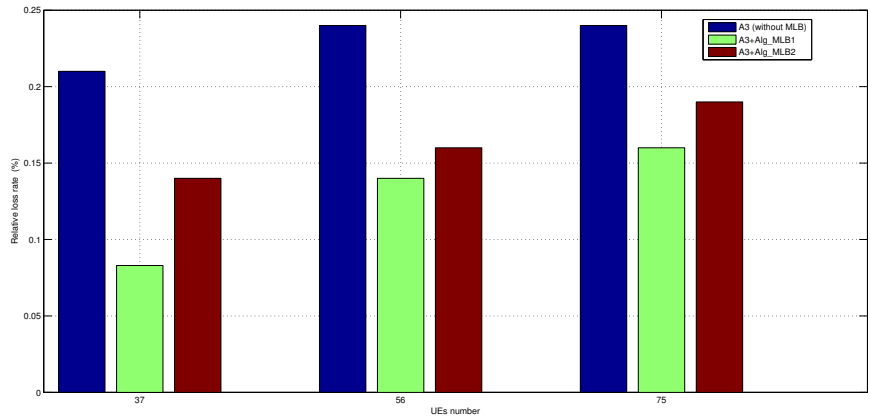

Figure. 7: Relative loss ratio vs UEs density

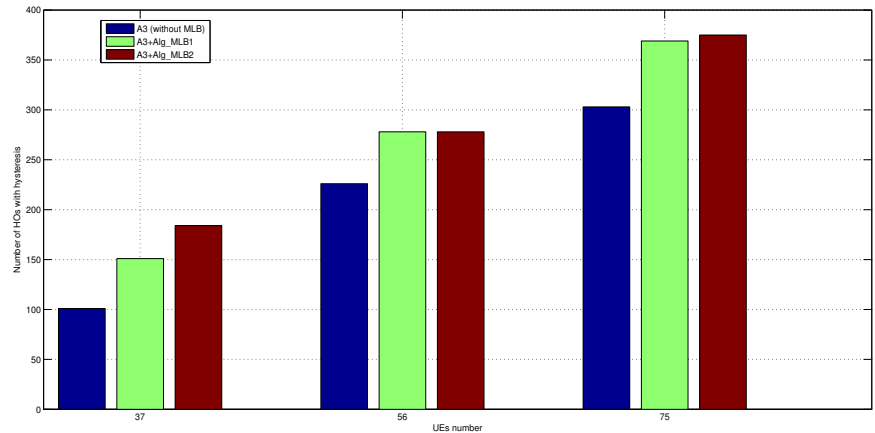

Figure. 8: Number of successful HO vs UEs density

The load is quantified through the number of active UEs. Notice that the number of HO is significantly important for both *Alg_MLB1* and *Alg_MLB2* when compared with *Alg_without_MLB*. This is explained by the fact that upon triggering MLB, a transfer of excess traffic in the overloaded cell to less loaded neighbouring cells is performed. This load transfer is achieved by adjusting the hysteresis value which favours handovers. We can also from Figure 8 establish a comparison between the performance of *Alg_MLB1* and *Alg_MLB2*. Note especially for relatively low loads that the number of handovers in the LTE network for the *Alg_MLB1* is considerably lower than for *Alg_MLB2*. As soon as the amount of available resources is beyond the threshold $Th_{AvailMLB}$ . This may be explained by the following arguments:

- For *Alg_MLB1*: There is a gradual and linear reduction of the hysteresis value (and thus increase of neighbouring cell size) as soon as the MLB process is initiated and the available load level of the neighbouring cell(s) is greater than $Th_{AvailMLB}$
- For *Alg_MLB2*: The hysteresis value of the neighbouring cell(s) decrease(s) sharply, promoting hence the number of handovers compared to *Alg_MLB1*.

The comparative study investigated above shows that *Alg_MLB1* algorithm provides indeed lower loss ratio when compared with *Alg_MLB2* however it may cause in return an increase of HO frequency.

## 5. CONCLUSION

In this article we implemented through ns-3 new load balancing algorithms at ENodeBs level for DL data transmission in LTE networks. We evaluated next the performance of these algorithms through appropriate scenarios to MLB context. Performance evaluation focused specifically on the investigation of the MLB impact on the global network throughput, the average throughput per sector and the loss ratio and the number of successful HO, for different UEs densities. The obtained results highlight comparative performance study between the simulation results with and without MLB. We show through these investigations the advantage of each MLB algorithm in finding convenient compromise in terms of throughput, loss ratio and number of successful HO. As perspectives, we plan to conduct more exhaustive simulations on joint-optimization of MLB and MRO. This leads to finding, for a given hysteresis value of a given cells' loads, the optimal

TTT value. We also plan in future to model MLB using the Markov decision process. This will allow to find the optimal values of various activation thresholds, acceptance and deactivation of MLB according to the load algorithm. Our main contribution in this paper is the implementation of MLB algorithms on ns-3. Note that the latest version on which we worked does not implement the MLB in the LTE module of ns-3.

## Authors

**Mohamed Escheikh** received in 1992 the Diploma degree in electrical engineering. In 1994, he received a Master's degree and in 2001 a PhD degree with Distinction all in electrical engineering from National Engineering School of Tunis ENIT (Tunisia). He is currently Assistant Professor at ENIT since 2001, member of SYSCOM laboratory at ENIT since 1992 and associate researcher with Vespa Team of Cedric CNAM Paris. His research interests include in particular dependability analysis, queuing systems, Petri nets, mobile networks optimization, cloud computing and power management.

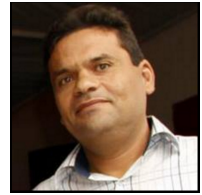

**Hana Jouini** received the Master's. degree in mathematics and computer science at the university of Paris Descartes, Paris, France in 2013. She is currently a Ph. D. student in computer science and telecommunication at the National engineering school of Tunis and at CNAM Paris.

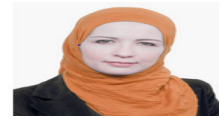

**Kamel Barkaoui** is full professor at the Department of Computer Science of Conservatoire National des Arts et Métiers (CNAM - Paris) since 2002. He holds a Ph.D in Computer Science (1988) and Habilitation à Diriger des Recherches (1998) from Université Paris 6 (UPMC). His research interests include formal methods for verification, control and performance evaluation of concurrent and distributed systems. He has published over 150 papers on these topics, as well as supervising 25 PhD theses and several MSc thesis projects. He received the 1995 IEEE Int. Conf. on System Man and Cybernetics Outstanding Paper Award. He served on PCs and as PC chair and OC chair of a number of international workshops and conferences in his areas of research. He is the SC chair of the International Workshop on Verification and Evaluation of Computer and Communication Systems (VECoS). He was recently general co-Chair of the 18th International Symposium on Formal Methods (2012) and general chair of 35th International Conference on Application and Theory of Petri Nets and Concurrency (Petri Nets 2014) and the 14th International Conference on Application of Concurrency to System Design (ACSD'2014). He is associated editor of the International Journal of Critical Computer-Based Systems (IJCCBS).

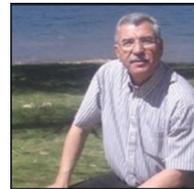